\documentstyle[prl,aps,epsf,multicol]{revtex}
\begin{document}
\draft
\title{Quasiparticles in the $111$ state and its compressible ancestors}
\author{M.~Y.~Veillette$^1$, L.~Balents$^1$, and Matthew P.A. Fisher$^2$}
\address{$^1$Physics Department, University of California, Santa
Barbara, CA 93106 \\
$^2$Institute for Theoretical Physics, University of California, Santa Barbara, CA 93106}
\date{Date:\today}
\maketitle

\begin{abstract}
  We investigate the relationship of the spontaneously inter-layer
  coherent ``$111$''state of quantum Hall bilayers at total filling factor
  $\nu=1$ to ``mutual'' composite fermions, in which vortices in one
  layer are bound to electrons in the other.  Pairing of the mutual
  composite fermions leads to the low-energy properties of the $111$
  state, as we explicitly demonstrate using field-theoretic
  techniques.  Interpreting this relationship as a {\sl mechanism} for 
  inter-layer coherence leads naturally to two candidate states with
  non-quantized Hall conductance: the mutual composite Fermi liquid,
  and an inter-layer coherent charge $e$ Wigner crystal.  The
  experimental behavior of the interlayer tunneling conductance and
  resistivity tensors are discussed for these states.
\end{abstract}

\begin{multicols}{2}
\narrowtext 

Double-layer quantum Hall systems exhibit a wealth of fascinating
behavior, among the most beautiful of which is the integer quantum
Hall effect at {\sl total} filling fraction
$\nu=1$.\cite{eisenstein92}\ In the absence of inter-layer tunneling
(which can be tuned to be arbitrarily small experimentally), the mere
existence of the quantum Hall effect at this filling factor is
possible only due to strong Coulomb interactions, which stabilize the
so-called $111$ state (see below).  As discussed theoretically by
Girvin {\sl et. al.}, this state may also be viewed as an easy-plane
pseudo-spin Quantum Hall FerroMagnet (QHFM), having spontaneously
broken the $U(1)$ symmetry corresponding to the conservation of the
charge difference between the two electron gases.  A remarkable
richness of behavior was predicted to arise in response to in-plane
magnetic fields and changes in temperature, much of which has indeed
been verified experimentally.

The $111$ state occurs for small inter-layer separation $d<d_c$, in
which the Coulomb interactions between electrons in opposite layers
are strongest.  Much less well understood is the behavior of bilayers
at $\nu=1$ for larger separations.  In the limit $d\rightarrow\infty$,
the layers become decoupled, and it is believed that each layer forms
an Independent compressible Composite Fermi Liquids\cite{HLR} (ICFL).
For intermediate $d>d_c$, the ground state is not known.  Moreover, in
this range such bilayers exhibit unexplained and somewhat puzzling
behavior.  Coulomb drag measurements show a surprising low-temperature
saturation of the trans-conductivity, quite different from the
predictions for the ICFL state.  Recent experimental measurements of
the nonlinear interlayer tunneling conductance show considerable
structure for $d>d_c$.  Surprisingly, most of this structure is
preserved for $d<d_c$, being modified only in a narrow range of low
voltage bias.

In this paper, we exploit the equivalence of the $111$ state to a
{\sl p-wave superconductor} (pSC) of {\sl Mutually Composite Fermions}
(MCFs), postulated earlier by Morinari\cite{Morinari}\ and
rediscovered recently by Kim {\sl et al}.\cite{Demler}\ These MCFs
themselves are similar to but distinct from the usual composite
fermions, and in particular in and of themselves already embody strong
inter-layer correlations.  We explore the MCF formulation in much more
detail than prior treatments, demonstrating that the Lagrangian of the
pSC state is {\sl dual} to the earlier FM picture.  The $111$ state
therefore provides an explicit realization of the $2+1$-dimensional
bosonization and duality formulation espoused in
Ref.~\onlinecite{NL3}.  Our results considerably deepen the
understanding of the charged and neutral sectors of the $111$ state,
and their coupling to (pseudo-)spin.  Finally, this analysis provides
two natural candidate ground states for $d\gtrsim d_c$: the Mutual
Composite Fermi Liquid (MCFL) of unpaired MCFs, and a charge $e$
Wigner crystal with coincident pseudo-spin superfluidity.  As an
experimental means of searching for these two possible intermediate
states, we investigate the corresponding resistivity tensors and the
interlayer tunneling conductances.  The MCF liquid has {\sl metallic}
intra-layer longitudinal resistivity and a constant finite {\sl Hall
  drag} at low temperatures, but a pseudo-gap in the inter-layer
tunneling conductance.  The charge $e$ Wigner crystal is {\sl
  insulating}, but should exhibit a {\sl sharp
  inter-layer tunneling conductance peak} at low temperatures due to
inter-layer phase coherence.

Simple algebraic manipulations of lowest Landau level wave functions
suggest a relation between the $111$ state and MCFs.  Similar
considerations have been successfully used to relate the $331$ state
to a pSC phase of ordinary composite Fermions.  The simplest
description of the QHFM is in terms of the $111$ ,
\begin{equation}
\Psi_{111} = \prod_{i<j} (z_i-z_j)(w_i-w_j)\prod_{ij}(z_i-w_j) \Psi_G,
\end{equation}
where $\Psi_G = \exp[ -\sum_i (|z_i|^2+|w_i|^2)/4\ell^2]$, and $\ell$
is the magnetic length, and $z,w = x+iy$.  Using the useful identity
$\prod_{i<j}[(z_i-z_j)(w_i-w_j)]/\prod_{ij}(z_i-w_j) = {\rm det}
[1/(z_i-w_j)]$, 
this can be rewritten as
\begin{equation}
  \Psi_{111} = \prod_{ij} (z_i-w_j)^2 {\rm det}[1/(z_i-w_j)] \Psi_G.
  \label{MCFpSC}
\end{equation}
The latter rewriting demonstrates that the $111$ wavefunction is the
product of a BCS pair wavefunction (the ${\rm det}[1/(z_i-w_j)]$
factor) and a phase-carrying factor $\Pi_{ij} (z_i-w_j)^2$.  This phase factor
can be interpreted in the usual quantum Hall fashion in terms of flux
attachment.  In particular, this factor is equivalent to attaching two
flux quanta (or more precisely vortices) of layer-1 flux to the
electrons in the second layer, and vice-versa.  Since an even number
of flux quanta are attached to each particle, the composite objects so
formed remain {\sl fermionic}.

We next turn to a field-theoretic formulation of this flux attachment.
Denoting the microscopic electron (annihilation) fields $c_\alpha({\bf
  x})$ (where $\alpha=\uparrow,\downarrow$ indexes the two pseudo-spin
components (layers)), we define MCF operators $\psi_\alpha({\bf x}) =
\exp[ i K_{\alpha\beta} \int\! d^2{\bf x'} \Theta({\bf x-x'})
n_\beta({\bf x'}) ] c_\alpha ({\bf x})$.  Here $\Theta({\bf x})$ is the angle of the
vector ${\bf x}$ in the plane (spatial (2D) vectors are indicated in
boldface), the matrix $K=2\sigma^1$ (We denote the 
Pauli matrices $\sigma^\mu = (\sigma^z,\sigma^x,\sigma^y)$ for
$\mu=0,1,2$), and $n_\alpha = c_\alpha^\dagger
c_\alpha^{\vphantom\dagger}$.  In terms of the $\psi$ variables,
standard techniques give the Euclidean Lagrange density for the system,
${\cal L} = {\cal L}_\psi + {\cal L}_a,$ with
\begin{eqnarray}
  {\cal L}_\psi & = & \overline{\psi}_\alpha(\partial_0 - \mu - i
  \tilde{a}_0^\alpha )\psi^{\vphantom\dagger}_\alpha  + {1 \over
    {2m^*}} \left|(\partial_j - i \tilde{a}_j^\alpha
    )\psi_\alpha\right|^2 \label{Lpsi} \\ 
  {\cal L}_a & = & {i \over
      {4\pi}}K^{-1}_{\alpha\beta} \epsilon^{\mu\nu\lambda}
    a_\mu^\alpha \partial_\nu a_\lambda^\beta ,    \label{chern1}
\end{eqnarray}
where Greek and Latin subscripts indicate $3$-vector and $2$-vectors,
respectively ($\partial_\mu = (\partial_\tau,\bbox{\nabla})$,
$\partial_i = \bbox{\nabla}$), $a_\mu^\alpha$ comprise a pair of
Chern-Simons (CS) gauge fields, and $\mu$ is the chemical potential
(usually taken positive).  We use the notation that a gauge field with
a tilde indicates the difference of CS and external gauge fields, e.g.
$\tilde{a}_\mu^\alpha = a_\mu^\alpha-A_\mu^\alpha$, where
$A_\mu^\alpha$ is an external gauge field used both to include the
magnetic field and for generating correlation functions by
differentiation.  At the mean-field level, $\langle
\tilde{a}_\mu^\alpha \rangle =0$ at $\nu=1$ -- we consider
fluctuations about this limit.  In Eq.~\ref{chern1}, we have dropped a
Coulomb interaction term which will turn out to be irrelevant for the
qualitative physics within the $111$ state -- it will be included when
we return to the unpaired MCF liquid below.  Instead, we assume for
the moment that the interactions between MCFs (from both Coulomb and
gauge sources) are such that they favor a
pSC\cite{Morinari,Bonesteel}.  The BCS reduced {\sl Hamiltonian}
contains the additional term
\begin{equation}
  H_\Delta = \int \! {{d^2\bbox{k}} \over (2\pi)^2} \left[
    \Delta_{\bbox{k}-\bbox{\tilde{a}}^s} \psi_{ \bbox{k}
      \uparrow}^\dagger \psi_{-\bbox{k} 
      \downarrow}^\dagger + {\rm h.c.}\right], \label{HDelta}
\end{equation}
where $\psi_{\bbox{k} \alpha} = \int\! d^2{\bbox{x}}
e^{i{\bbox{k}\cdot \bbox{x}}}\psi_\alpha(\bbox{x})$.  In a pSC (more
precisely, an $L^z=0$ triplet state) the pair field $\Delta_{\bbox{k}}
= e^{i\varphi} v(k^2) (k_x+ik_y)$, where $\varphi$ is the phase of the
pair wavefunction, and $v(k^2)$ is a smooth function of its argument.
For simplicity, we will take $v(k^2) = v$ constant, adequate for {\sl
  universal} properties.  In the wavevector-dependent gap in
Eq.~\ref{HDelta}, we have made the replacement $\bbox{k} \rightarrow
\bbox{k} - \bbox{\tilde{a}}^s$, with $a_\mu^{c/s} \equiv
(a_\mu^\uparrow \pm a_\mu^\downarrow)/2$ (and similarly for
$A_\mu,\tilde{a}_\mu$).  This inclusion of gauge fields is justified
by more careful microscopic calculation, and physically reflects the
fact that the $z$'s and $w$'s in Eq.~\ref{MCFpSC}\ are {\sl electron}
(not MCF) coordinates.  

We then perform the key step of a combined
particle-hole transformation and phase rotation
\begin{equation}
  \Psi_\uparrow = e^{-i\varphi/2}\psi_\uparrow, \qquad
  \Psi_\downarrow = e^{i\varphi/2}\psi^\dagger_\downarrow.
  \label{doublevalued}
\end{equation}
The phase rotation in Eq.~\ref{doublevalued}\ apparently
``neutralizes'' the $\Psi$ fermions.  Note, however, that the
transformation {\sl becomes double valued in the presence of $\pm
  2\pi$ vortices in $\varphi$}, which has important consequences to
which we will return shortly.  Eq.~\ref{HDelta}\ can then be
re-expressed in Lagrangian form.  Combining it with
Eqs.~\ref{Lpsi}-\ref{chern1}\ gives ${\cal L} = {\cal L}_\Psi + {\cal
  L}_a + {\cal L}_{\rm irr.}$, with 
\begin{eqnarray}
  {\cal L}_\Psi & = & \overline\Psi \big[\partial_0\! -\!
  i\tilde{a}_0^s \! + \! iv\bbox{\sigma}\cdot(\bbox{\nabla}\! -\!
  i\bbox{\tilde{a}}^s) \! -\! \mu\sigma^z\big]\Psi,
   \label{LDelta} \\ 
  {\cal L}_{ac} & = & {i \over {4\pi}}\epsilon^{\mu\nu\lambda}
  a_\mu^c \partial_\nu a_\lambda^c, \qquad  {\cal L}_{as} =   - {i \over {4\pi}}\epsilon^{\mu\nu\lambda}
  a_\mu^s \partial_\nu  
    a_\lambda^s ,\label{La} \\
  {\cal L}_{\rm irr.} & = & (\partial_\mu\varphi -2\tilde{a}_\mu^c)
  {\cal J}^\mu +  
  {1 \over {8m^*}}|\bbox\nabla\varphi-2\bbox{\tilde{a}}^c|^2 \overline\Psi
  \sigma^0 \Psi \nonumber \\
  & & - {1 \over {2m^*}}\overline\Psi \sigma^0
  (\bbox\nabla-i\bbox{\tilde{a}}^s)^2\Psi. \label{Lirr}
\end{eqnarray}
Here ${\cal L}_a = {\cal L}_{ac}+{\cal L}_{as}$ and ${\cal
  J}^0 = \overline\Psi i\sigma^0\Psi/2$, $\bbox{{\cal J}} =
i[\overline\Psi(\bbox{\nabla}-i\bbox{\tilde{a}}^\sigma) \Psi - (\bbox{\nabla}+i\bbox{\tilde{a}}^\sigma) \overline\Psi \Psi]/2m^*$.  


An effective field theory is obtained by ``coarse graining'' --
i.e. integrating out gapped fermion modes at large ${\bf k}$.  This
reduces the ultraviolet momentum cut-off to
$\Lambda$ and also generates a ``kinetic'' term for the charge sector,
\begin{equation}
  {\cal L}_\varphi = {n_{SF} \over {2m v^2}}(\partial_0\varphi -
  2\tilde{a}_0^c)^2 + {n_{SF} \over {2m}} (\bbox\nabla\varphi -
  2\bbox{\tilde{a}}^c )^2. \label{Lphi}
\end{equation}
In the theory with the reduced cut-off, power-counting can be applied.
The terms in ${\cal L}_{\rm irr.}$ (Eq.~\ref{Lirr}) are {\sl
  irrelevant}, and will thus be 
neglected hereafter (though they could  give rise to some effects at
non-zero temperature and frequency, as do the very similar
Doppler shift couplings in d-wave superconductors).

Doing so, the Lagrangian becomes explicitly spin-charge separated!
The charge sector is governed by ${\cal L}_c = {\cal L}_\varphi +
{\cal L}_{ac}$, offering a physical interpretation as charge $2e$ {\sl
  composite bosons} (Cooper pairs) at an effective filling factor
$\nu_{\rm eff} = 1/4$.  With the ($4\times 4$) conductivity tensor in
the usual basis defined by $E_i^\alpha =
\rho^{\alpha\beta}_{ij}J^\beta_j$, it is convenient to introduce
charge and spin conductivities,
$\sigma_{ij}^{c/s}=2(\sigma_{ij}^{\uparrow\uparrow}\pm \sigma_{ij}^{\uparrow\downarrow})$.  The
(charge) Hall conductivity is quantized to $\sigma^c_{xy}=e^2/h =
1/2\pi$ (in our units), as seen by choosing the gauge $\varphi=0$ and
integrating out $a_\mu^c$ to obtain a CS term for $A_\mu^c$.  The spin
sector Lagrangian is ${\cal L}_s = {\cal L}_\Psi + {\cal
  L}_{as}$, describing massive Dirac Fermions coupled to a spin
CS gauge field.  To analyze ${\cal L}_s$, we integrate out the
$\Psi$ fermions, which generates
for $\mu>0$ a CS term and nominally irrelevant Maxwell and
higher-order in gradient corrections for $\tilde{a}^s_\mu$.
Therefore the {\sl effective} Lagrangian for the spin sector, ${\cal
  L}_s \rightarrow {\cal L}_{s}^{\rm eff.}$, is
\begin{equation}
  {\cal L}_{s}^{\rm eff.} = {i \over
    {4\pi}}\epsilon^{\mu\nu\lambda} \tilde{a}_\mu^s\partial_\nu
  \tilde{a}_\lambda^s + {1 \over
    {2\lambda}}(\tilde{e}_j^2-\tilde{b}^2)+ O[\partial^3\tilde{a}^2] +
  {\cal L}_{as},  \label{Lseff0} 
\end{equation}
where $\lambda \sim \pi \mu/v^2$ is a non-universal ``dielectric''
constant, and $\tilde{e}_j = v^{-1}(\partial_j \tilde{a}^s_0 - \partial_0
\tilde{a}^s_j)$, $\tilde{b}= \epsilon_{ij}\partial_i\tilde{a}^s_j$.
Significantly, the coefficient of $1/4\pi$ in the CS term above is a
factor of two {\sl larger} than what might naively be expected from
the massive Dirac fermion in Eq.~\ref{LDelta}.  To obtain the correct
value, it is necessary to take into account the proper boundary
conditions ($\mu \rightarrow -\infty$) on the theory, which yields
an edge state for $\mu>0$.  Alternatively, the CS term may be
obtained by considering a more general pair wavefunction $\Delta_{\bf
  k} = v(k_x + i \varepsilon k_y)$ with arbitrary $\varepsilon$.
Perturbing the non-chiral (and hence without a self-generated CS term)
$k_x$ state with small $\varepsilon$ correctly gives the CS content of
the chiral state, which persists unaltered due to its topological
nature as the state is adiabatically continued from $\varepsilon=0^+$
to $\varepsilon=1$.

Note that upon combining all the terms in ${\cal
  L}_s^{\rm eff.}$, there is a cancellation of CS contributions
for the fluctuating field $a_\mu^s$ (but not the external field
$A_\mu^s$).  Now integrating out $a_\mu^s$ gives
\begin{equation}
  {\cal L}_s^{\rm eff.} = {n_{\scriptscriptstyle SF}^s \over 2}
  \left[(A_j^s)^2\!+\! v^{-2}(A_0^s)^2\right]  +
  {{i\sigma_{xy}^s} \over {2}}\epsilon^{\mu\nu\lambda}
  A_\mu^s \partial_\nu 
  A_\lambda^s,  \label{Lseff}
\end{equation}
where $n_{\scriptscriptstyle SF}^s = \lambda v^2/4\pi^2$ is a {\sl spin superfluid
  density} (stiffness), so that this state is a pseudo-spin
superconductor -- the QHFM!  Interestingly, this state also exhibits a
hidden {\sl spin Quantum Hall effect}.  The {\sl spin Hall
  conductance} from Eq.~\ref{Lseff} is $\sigma_{xy}^s \neq \frac{\hbar^2}{h}=\frac{\hbar}{2 \pi}  = 1/2\pi$, the lack of quantization of $\sigma_{xy}^s$
being due to corrections from the non-universal $O[\partial^3
\tilde{a}^2]$ terms in Eq.~\ref{Lseff0}.  The non-universality of
$\sigma_{xy}^s$ is perhaps natural since the $U(1)$ symmetry
generated by $S^z$ is spontaneously broken.

Next consider the quasiparticle structure.  For charge $2e$ bosons at
$\nu_{\rm eff}=1/4$, the quasiparticle excitations, which correspond
to the (smallest) $2\pi$ vortices in $\varphi$, carry the charge
$\nu_{\rm eff.} \times 2e = e/2$, as can also be deduced directly from
Eq.~\ref{Lphi}.  Remarkably, owing to the implicit coupling in
Eq.~\ref{doublevalued}, this excitation also carries spin!  In
particular, the $\Psi$ fermions experience a {\sl cut} ($\pi$ flux)
upon encircling the vortex.  Because the XY spin operator $S^+ \sim
\Psi^\uparrow \Psi^\downarrow$ is bilinear in fermions, the charge
$e/2$ quasiparticle is thus tied to a $2\pi$ spin-flux vortex (in
$S^+$) (see also below).  Moreover, because of the spin quantum Hall
conductivity, this flux induces a {\sl non-universal} moment $\langle
S^z\rangle = \pm \pi\sigma_{xy}^s$.  We identify this excitation with
the {\sl meron} of Ref.~\cite{yang_moon}\ (the moment arises in that
picture from pseudo-spin canting in the meron's core).  Even-flux
vortices in $\varphi$ and $S^+$ leave Eq.~\ref{doublevalued}\ 
single-valued, and remain spin-charge separated.  Due to the Higgs
phenomena, the $\pm 4\pi$ vortices in $\varphi$ are 
screened, cost finite energy, and therefore unbind at any finite
temperature, giving activated contributions to the Hall effect.  The
merons, however, are tied to spin vortices which interact
logarithmically and therefore exhibit a Kosterlitz-Thouless transition
at finite temperature.

We now turn to the connection of the above formalism to the pseudospin
magnetization approach of Refs.~\onlinecite{yang_moon}.  To do so, we
return to ${\cal L}_s = {\cal L}_d+{\cal L}_{as}$
(Eqs.~\ref{LDelta},\ref{La}).  Following the reasoning of
Ref.~\onlinecite{NL3}, we argue that the CS gauge field ``bosonizes''
the Dirac fermions into relativistic charged bosons.  As in the much
more established 1+1-dimensional bosonization mapping, the expressions
for currents in terms of bosons are much simpler than those
for the fermion fields, and we are presently unable to derive the
latter.  Instead, we will determine the form of the ``bosonized''
Lagrangian by requiring that it produce the same generating function
for current--current correlators.  Thus we seek an equivalent
representation for the partition function
\begin{equation}
  {\cal Z}_\sigma[A_\mu^s] = \int [d\overline\Psi
  d\Psi][da_\mu^s] e^{- \int\! d^3x_\mu {\cal
      L}_s} = e^{- \int\! d^3x_\mu {\cal L}_s^{\rm
      eff.}[A_\mu^s]} .
\end{equation}
Referring back to Eq.~\ref{Lseff}, we recognize that the first term in 
${\cal L}_s^{\rm eff.}$ is readily obtained from the usual
``Higgs'' mechanism if $A_\mu^s$ is minimally coupled to a U(1)
boson which condenses.  To reproduce the second (Chern-Simons) term in
${\cal L}_s^{\rm eff.}$ we introduce in addition a massive Dirac
field which also carries the U(1) ``charge'' (actually spin).  Thus
\begin{equation}
  {\cal Z}_\sigma[A_\mu^s] = \int [d\overline\eta d\eta][d\theta] 
  e^{-\int d^3x_\mu \, {\cal L}_s^{\rm dual}},
\end{equation}  
where
\begin{eqnarray}
  {\cal L}_s^{\rm dual} & = & {n_{\scriptscriptstyle SF}^s \over
    {2mv^2}}(\partial_0 \theta - 2A^s_0)^2 +
  {n_{\scriptscriptstyle SF}^s \over {2m}}| \bbox{\nabla}\theta - 2{\bf A}^s|^2
  \nonumber  \\
  & & + \overline\eta\left[\partial_0\! -\!
    iA_0^s \! + \! iv\bbox{\sigma}\cdot(\bbox{\nabla}\! -\!
    i\bbox{A}^s) \! -\! M\sigma^z \right]\eta \nonumber \\
  & & + \gamma \left[ e^{i\theta} \overline{\eta} \sigma^y
    \overline{\eta} + {\rm h.c.} \right]. \label{Ldual}
\end{eqnarray}
From Eq.~\ref{Ldual}, we identify $S^{+} \sim e^{i \theta}$. In
Eq.~\ref{Ldual}, in addition to a Dirac Lagrangian of the usual form,
we have also included an ``anomalous'' coupling which exchanges the
spin between the $\theta$ boson and $\eta$ fermions.  Given only the
single physical U(1) spin-rotation symmetry, such a coupling is
allowed and indeed is required to reproduce the {\sl un-quantized}
spin Hall conductivity in Eq.~\ref{Lseff}.

Having established and explored the equivalence of the paired MCF
state and the QHFM, we now turn to a discussion of possible ``quantum
disordered'' ground states suggested by this work.  Specifically, we
consider cases in which the (charge) Hall resistance is {\sl
  unquantized}, motivated by the experimental observation of poorly
developed Hall plateaus.  These phases can be described loosely by the
proliferation {\sl in the ground state} of charge vortices, i.e. point
defects around which $\oint \vec\nabla\varphi \cdot d\vec{r} = 2\pi
N$, with integer $N$.  The two possible phases of interest correspond to the cases
in which $(i)$ only vortices with even $N$ proliferate, leaving
$\theta$ single-valued, and $(ii)$ all vortices are unbound.

The ``doubly quantized'' vortices in case $(i)$ are conventional,
insofar as they leave the singular gauge transformation in
Eq.~\ref{doublevalued}\ single-valued.  Hence these vortices interact
only weakly with the other excitations of the system.  Their
proliferation can thus be analyzed using conventional methods.  In
particular, by performing a $2+1$-dimensional duality transformation
on the XY-model in Eq.~\ref{Lphi}, the proliferated state can be
described as a condensate of pairs (due to the even $N$ condition) of
vortices (merons).  (Dual) Phase coherence of the vortex-pair
wavefunction implies the quantization of charge in units of half the
composite boson charge, or $e$.  Thus the bilayer charge density of
$e$ per area $\ell^2$ is distributed into a Wigner crystal of charge
$e$ per unit cell.  This {\sl charge $e$ Wigner crystal} is of course
an electrical insulator (with $\sigma^c_{xx}, \sigma^c_{xy}
\rightarrow 0$ as $T\rightarrow 0$ provided the sliding mode is even
infinitesimally pinned).  Because the {\sl paired} vortex condensate
respects spin-charge separation, however, interlayer phase coherence
(spin superfluidity) is maintained.  An alternative picture for this
phase is as a staggered bilayer crystal in which vacancy-interstitial
pairs made from opposite layers have Bose condensed.  Thus the {\sl
  (pseudo)spin} conductances are very different:
$\sigma_{xx}^{s}(\omega) = n_{\scriptscriptstyle SF}^s/i\omega$, so
that the zero-bias spin conductivity is {\sl infinite}, while
$\sigma_{xy}^s$ is a non-universal constant.  Note that this implies
$\rho^s \rightarrow 0$ for $T\rightarrow 0$, so the single-layer
resistivity $\rho^{\uparrow\uparrow} \approx \rho^c$ does not manifest
superfluidity.  At strictly zero temperature, $\rho^c_{xx} =\infty$,
so that the drag resistance would {\sl diverge}.  For $T>0$, it is
natural to expect a peak in $\rho_{xx}^{\uparrow\uparrow}$ as a
function of $d/\ell$ as a precursor effect.  Because of inter-layer
phase coherence, however, the charge $e$ Wigner crystal should
exhibit a zero-bias tunneling conductance peak as in the QHFM.

In case $(ii)$, by contrast, the strong statistical interaction of
individual merons amongst themselves and with other excitations
renders their proliferation a strong-coupling problem.  On physical
grounds, however, we speculate that their presence in the ground state
simply destroys all effects of MCF pairing on long length and time
scales.  We are therefore led to consider a simple model of MCFs {\sl
  without pairing}, analogous (but distinct from) the composite Fermi
liquid.  This MCF Liquid is described simply by the Lagrangian $L =
\int\! d^2\vec{x} [{\cal L}_\psi + {\cal L}_a] + L_C$, where ${\cal
  L}_\psi$ and ${\cal L}_a$ are given in Eqs.~\ref{Lpsi},\ref{chern1},
and due to the non-vanishing MCF compressibility, it is necessary a
priori to include an additional long-range Coulomb interaction term
$L_C$  (see Refs.~\onlinecite{Kim,Sakhi,Stern}).  


Drag between {\sl weakly-coupled} layers in nearly-independent
$\nu=1/2$ composite Fermi liquid states has been considered previously
by many authors.\cite{Kim,Stern}\ In that case, the drag {\sl
  resistivity} $\rho^{\uparrow\downarrow}$ is truly perturbative in
the interlayer interaction., and the formalism due originally to
Zheng+MacDonald\cite{ZhengMacDonald}\ can be applied to yield a small
{\sl longitudinal drag resistivity} $\rho_{xx}^{\uparrow\downarrow}
\sim T^{4/3}$ at low temperature.\cite{Stern}\ The inherently strong
inter-layer interactions in the MCF liquid unfortunately preclude this
approach, and one is reduced to a diagrammatic treatment as in
Refs.~\onlinecite{Kim,Sakhi}.  This diagrammatic treatment is much
less satisfactory, but reasoning along the lines of
Refs.~\onlinecite{Kim,Sakhi}\ suggests, and we therefore propose, that
$\rho_{xx}^{\uparrow\downarrow} \sim T^{4/3}$ also obtains for the MCF
liquid.  Unlike, the ICFL, however, the Random Phase
Approximation\cite{HLR}\ already gives a non-zero ``Hall drag''
$\rho_{xy}^{\uparrow\downarrow} = 4\pi$, and we expect this is robust.
Thus the longitudinal drag resistivity is small also in this case, and
only the Hall drag is expected to deviate substantially from the ICFL
limit.  Furthermore, the interlayer tunneling conductance in the MCF
liquid, like that of the ICFL state, is expected to exhibit a
pseudo-gap due to orthogonality catastrophe and poorly-screened
Coulombic effects.

L.B. and M.V. were supported by the NSF--DMR--9985255 and the Sloan
and Packard foundations.  M.P.A.F. was supported by PHY--9907949,
NSF--DMR--9704005.

\end{multicols}
\end{document}